\documentclass[aps,onecolumn,10pt]{revtex4-2}
\usepackage{amsmath}
\usepackage{graphicx}
\usepackage{xcolor}
\usepackage{hyperref}
\usepackage{amssymb}
\usepackage{ulem} 

\usepackage{geometry}
\DeclareUnicodeCharacter{2212}{-}

\begin{document}

\title{Wind-wave growth over a viscous liquid}
\author{J. Zhang$^{1,2}$}
\author{A. Hector$^1$}
\author{M. Rabaud$^1$}
\author{F. Moisy$^1$}
\email{frederic.moisy@universite-paris-saclay.fr}
\affiliation{$^1$Universit\'e Paris-Saclay, CNRS, FAST, F-91405 Orsay, France. \\
$^2$PMMH, CNRS, ESPCI Paris, Universit\'e PSL, Sorbonne Universit\'e, Universit\'e Paris-Cit\'e, F-75005, Paris, France.}

\date{\today}

\begin{abstract}
\vspace{1.7 cm}

Experimental and theoretical studies on wind-wave generation have focused primarily on the air-water interface, where viscous effects are small. Here we characterize the influence of the liquid viscosity on the growth of mechanically generated waves. In our experiment, wind is blowing over a layer of silicon oil, of viscosity 20 and 50 times that of water, and waves of small amplitude are excited by an immersed wave-maker. We measure the spatial evolution of the wave slope envelope using Free-Surface Synthetic Schlieren, a refraction-based optical method. Through spatiotemporal band-pass filtering of the surface slope, we selectively determine the spatial growth rate for each forcing frequency, even when the forced wave is damped and coexists with naturally amplified waves at other frequencies. Systematic measurements of the growth rate for various wind velocities and wave frequencies are obtained, enabling precise determination of the marginal stability curve and the onset of wave growth. We show that Miles' model, which is commonly applied to water waves, offers a reasonable description of the growth rate for more viscous liquids. We finally discuss the scaling of the growth rate of the most amplified wave and the critical friction velocity with the liquid viscosity.

\vspace{1.3 cm}

\end{abstract}

\maketitle

\section{Introduction} \label{sec:intro}

The generation of waves by the wind, whether on a puddle, a pond, or at sea, remains a fascinating phenomenon that has challenged researchers for decades~\cite{Janssen_2004,sullivan2010dynamics,Rabaud_2020,pizzo2021,ayet2022}.
Gaining a comprehensive understanding of this fundamental problem is of first importance for wave forecasting, assessing heat and mass transfers in the ocean, and in engineering applications involving liquid and gas transport in pipes or heat exchangers~\cite{boomkamp1996classification}.

Since the pioneering works of Phillips~\cite{phillips1957generation} and Miles~\cite{Miles_1957}, several decades of research have led to the following two-stage scenario. In the first stage, incoherent wrinkles of very low amplitude are excited by the turbulent pressure fluctuations of the wind (Phillips mechanism), resulting in a {\it linear} increase in wave energy over time~\cite{phillips1957generation,zavadsky2017water,li_shen_2022}, with a saturation to a finite amplitude governed by the liquid viscosity for moderate wind~\cite{Paquier_2016,Perrard2019,Nove_2020}. In the second stage, for sufficient wind, the air-water shear flow becomes unstable, leading to a subsequent {\it exponential} increase of the wave energy (Miles mechanism)~\cite{Miles_1957,plant1977growth,shemer2021spatially}. However, this highly idealized scenario still requires experimental and numerical confirmations.
In particular, there is no consensus on the nature of the transition between the Phillips' and Miles' regimes and its dependence on the various physical parameters of the problem~\cite{Perrard2019,li_shen_2022}. In this context, accurate laboratory measurements of growth rates under controlled conditions, for physical parameters distinct from the usual air-water configuration, are highly valuable.

The difficulty in modeling the wind-wave generation problem has several origins. First, the flow on the air side is turbulent, and both the characteristics of the mean velocity profile and the statistical properties of the turbulent stress fluctuations play significant roles. Second, because of the transport of wave energy, the {\it temporal} growth of waves translates into {\it spatial} growth~\cite{plant1977growth}, but a quantitative relationship between the two requires steady forcing conditions over long distances and durations, which are rarely satisfied in open sea conditions. Third, the wind forces  not only  waves but also currents, which may reach saturation on a timescale distinct from the growth time of the waves~\cite{zavadsky2017water}. Fourth, surface contamination, which is inevitable in open sea environments and large water tanks, can significantly impact the onset of wave generation~\cite{ryan2010recurrent}. Lastly, even in controlled laboratory experiments, the presence of walls introduces additional damping effects and undesirable wave reflections~\cite{hidy1966wind, plant1977growth, Caulliez_1998}.

In this paper, we consider the problem of wind-wave generation on viscous liquids (silicon oils of viscosity 20 and 50 times greater than that of water), with the aim of circumventing some of the previous difficulties: the flow in the liquid is purely laminar, enabling accurate analytical modeling; the surface drift remains moderate, simplifying the air-liquid energy transfer problem; the impact of surface contamination is reduced by the use of a liquid of low surface tension; and interference with reflected waves is reduced by the strong viscous damping.

As long as the liquid viscosity is not too high (typically less than 100~mm$^2$~s$^{-1}$), the two-stage scenario of the air-water configuration, with linearly increasing wrinkles at small wind (Phillips mechanism) and exponentially growing waves at large wind (Miles mechanism), is still relevant; larger liquid viscosities are not considered here, as they exhibit a very distinct behavior governed by the Kelvin-Helmholtz instability~\cite{Francis_1954,Miles1959generation,aulnette2019wind,Rabaud_2020,aulnette2022kelvin}. While previous studies using viscous liquids focused on waves naturally generated by the wind~\cite{Paquier_2015,Paquier_2016}, here we are interested in waves triggered by a small mechanical disturbance. Provided that their initial amplitude is sufficient, this forcing bypasses the first stage of Phillips~\cite{phillips1957generation}, and the system directly enters the second stage of Miles~\cite{Miles_1957}. This shortcut in the two-stage wave generation problem has the advantage of directly assessing the growth rate of each wave number, at the cost of disregarding the realistic conditions required for the ``true'' wave generation process starting from an initially undisturbed surface. As a result, different critical wind velocities may be expected for natural and mechanically generated waves.

In the exponential growth regime, Miles demonstrated that the normalized temporal growth rate of a given wave number $k$ is proportional to the normalized wind shear stress $(u^*/c)^2$, where $u^*$ is the friction velocity and $c(k)$ the phase velocity, with a coefficient $\beta$ that depends on the curvature of the mean velocity profile at the ($k$-dependent) critical height where the mean flow velocity matches the phase velocity. This growth is mitigated by the dissipation in the liquid, either laminar or turbulent, which defines in principle a critical friction velocity for the growth of this wave number $k$. However, unlike standard hydrodynamic instabilities arising from the infinitesimal disturbance of a base state, the two-stage transition here also depends on the incoherent base state from which the waves grow, which itself depends on the liquid viscosity and the wind forcing conditions (intensity, duration)~\cite{Perrard2019}. This complexity likely contributes to the persistent challenge in defining a clear critical wind velocity for the onset of wave generation under realistic conditions.

The growth rate of wind-generated waves has been the subject of a large number of experimental works, either from their temporal growth at a fixed point~\cite{larson1975wind, Kawai1979generation, geva2022excitation}, or from their spatial growth in steady conditions~\cite{hidy1966wind,Wilson_1973, mitsuyasu1982wind, grass2001measurement, tsai2005spatial, Liberzon_2011} (the equivalence between these two approaches is discussed in Ref.~\cite{plant1977growth}), but only in the air-water configuration. These measurements, initially limited to one-point probes, have made significant progress with the advent of optical methods resolved in space and time~\cite{Moisy09,Paquier_2015,Paquier_2016,shemer2019,yousefi2020,liu2022} and direct numerical simulations~\cite{lin2008direct,zonta2015growth,li_shen_2022,Wu_2021,wu2022revisiting,burdairon2023}, revisiting this old problem with new, high quality data. 

In our experiments, the spatial evolution of mechanically generated waves are measured using Free-Surface Synthetic Schlieren~\cite{Moisy09}, a refraction-based optical method offering a wave slope resolution of $3 \times 10^{-3}$. To isolate the spatial evolution of the wave at each forcing frequency, we use a dedicated spatiotemporal band-pass filtering. This method enables us to measure the growth or damping rate for each forcing frequency, even for cases above wind onset where natural waves are also present, from which we determine the marginal stability curve of the instability with unprecedented resolution. We show that Miles’ model, which is commonly applied to water waves, offers a reasonable description of the growth rate for more viscous liquids, and we propose a scaling for the growth rate of the most amplified wave with the liquid viscosity and friction velocity. We finally discuss the difference between the critical friction velocity that triggers wave growth in the case of mechanically generated waves and natural waves.

\section{Wave growth and attenuation mechanisms}
\label{Sec_growth_attenuation}

The spatial evolution of wind-generated waves results from a combination of energy transferred from the air flow and dissipated in the liquid. We first recall the modeling of these two contributions. We start from a simple harmonic surface deformation, $\zeta\left ( x,t \right ) = \zeta_0e^{i\left ( kx-\omega t \right)}$, of wave number $k$ and angular frequency $\omega$, propagating over a liquid layer of depth $h$, density $\rho_\ell$, and surface tension $\sigma$. In the absence of dissipation and forcing, $\omega$ and $k$ are real and are related by the dispersion relation
\begin{equation}
        \omega^2=\left ( gk+\frac{\sigma k^3}{\rho_\ell} \right) \tanh \left( kh \right)=gk\left ( 1+\frac{k^2}{k_{cap}^2} \right) \tanh \left( kh \right),
\label{eq:SDR}
\end{equation}
where $k_{cap}=2\pi/\lambda_{cap}=\sqrt{{\rho_\ell g}/{\sigma}}$ is the capillary wave number. The energy density per unit surface of the wave writes $E = \rho_\ell g \zeta_0^2 [ 1+(k/k_{cap})^2]$. Neglecting nonlinear interactions and surface currents, the transport and growth of the wave energy are governed by~\cite{Kahma_1988,grare2013growth,PEIRSON:2008}
\begin{equation}
\left(\frac{\partial}{\partial t} + c_g \frac{\partial}{\partial x} \right) E = P - D,
\label{eq:EPD}
\end{equation}
with $c_g = \partial \omega / \partial k$ the group velocity, $P$ the local power injected by the wind and $D$ the local power dissipated in the liquid by viscosity.

The injected power is the work per unit time of the wind stress (pressure and viscous shear stress) applied on the liquid surface, and can be modeled in two ways, depending on the wave amplitude. For very low wave amplitude, typically smaller than a fraction of the viscous sublayer thickness, the wavy shape of the liquid surface does not modify the flow in the air. In this case, only the turbulent pressure fluctuations in the air deform the interface, resulting in an injected power $P$ independent of the wave energy $E$, and hence a {\it linear} growth of $E$: this is the starting hypothesis of the Phillips model~\cite{phillips1957generation}. On the other hand, for larger wave amplitude, a feedback of the wave on the air flow takes place, and the wind stress becomes modulated by the wave profile. The resulting injected power $P$ is now proportional to $E$, yielding an {\it exponential} growth of the wave energy: this is the Miles model~\cite{Miles_1957}.

In the present paper, we are interested in mechanically generated waves of initial amplitude larger than the viscous sublayer thickness, which places us directly in the second stage of wave growth. In this regime, both $P$ and $D$ are proportional to $E$, so the right-hand-side of Eq.~(\ref{eq:EPD}) can be written in the form $P-D = \gamma E$, with $\gamma$ the temporal growth rate. In this framework, the onset of wave growth is defined by $\gamma=0$, when the injected power overcomes the viscous dissipation. 

We first focus on the injected power $P$. Since the flow in the air is necessarily turbulent in our problem, the wave-induced wind stress variations (pressure and viscous shear stress) are conveniently scaled with the applied mean shear stress $\tau = \rho_a u^{*2}$, with $u^*$ the friction velocity and $\rho_a$ the air density. In a developing boundary layer, this friction velocity $u^*$ depends on the free-stream wind $U_a$ and the fetch $x$ (distance along which the stress is applied)~\cite{Schlichting}.

To first order in wave slope, since the velocity of the surface is essentially vertical, the work per unit time of the horizontal component of the stress (shear stress) can be neglected, and only the vertical component (pressure) contributes to the injected power. This contribution writes $P = \Re \langle p^\dagger \dot \zeta \rangle$, where $ p^\dagger$ is the complex conjugate of the oscillating component of the pressure along the surface and $\dot \zeta$ the vertical velocity of the surface~\cite{Miles_1957,Janssen_2004}. It therefore depends on the phase relation between the surface elevation and the wave-induced pressure oscillation. Following Miles~\cite{Miles_1957}, we can write the oscillating pressure in the form $p=(\alpha + i \beta) \rho_a u^{*2} k \zeta$ to first order in wave slope, with $\alpha + i \beta$ a nondimensional parameter coding for this phase shift~\cite{wu2022revisiting}. Only the pressure contribution in phase with the wave slope ($\beta \neq 0$) provides work, yielding a growth rate $\gamma = P / E$ in the form~\cite{Miles_1957,plant1982relationship}
\begin{equation}
   \gamma = \beta \omega \frac{\rho_a}{\rho_\ell} \left( \frac{u^*}{c} \right)^2.
\label{eq:Miles} 
\end{equation}

The normalized phase velocity $c/u^*$ is often referred to as the ``wave age'', because young waves are short and have small phase velocity $c$. The key result of Miles theory~\cite{Miles_1957,Lighthill_1962} is that $\beta$ is proportional to the curvature $-U''(z_c)$, with $z_c$ the elevation of the critical layer where the wind velocity $U(z_c)$ matches the phase velocity of the considered wave. This indicates that too small or too large wavelengths, for which $z_c$ is such that $U''(z_c)$ vanishes (in the viscous sublayer or outside the boundary layer), cannot be amplified by this mechanism. The main difficulty in the determination of $\beta$ therefore relies in the detailed knowledge of the velocity profile in the air.

Since the original work of Miles, the experimental verification of the scaling (\ref{eq:Miles}), and the question of the extent to which the parameter $\beta$ can be considered a constant, have been the subject of numerous studies. For naturally growing waves, $\beta$ corresponds to the most amplified wave number, but it is possible to examine its dependence with $k$ using mechanically generated waves. Compilation of field and laboratory experiments suggests that the scaling $(u^*/c)^2$ approximately holds for the air-water interface over a significant range of wave ages $(1<c/u^*<20)$, with an average value $\beta \simeq 32\pm{16}$~\cite{plant1982relationship,Janssen_2004}. This large variability can be ascribed to a number of parameters, such as the dependence with wave number, the presence of currents, the incorrect modeling of wave dissipation for nonlaminar flow in the liquid, nonlinear effects (dependence with wave slope~\cite{PEIRSON:2008}), etc.

We now describe the dissipation in the liquid, focusing on the laminar dissipation mechanisms, which are relevant for the viscous liquids considered here. Viscosity introduces a negative imaginary part in the dispersion relation, $\omega = \omega_r + i \omega_i$, yielding a temporal damping rate $\gamma = -D/E = 2\omega_i < 0$ of the wave energy. The full dispersion relation for arbitrary viscosity (without side effects) is given by Lamb~\cite{Lamb} in infinite depth, later improved by LeBlond and Mainardi~\cite{Leblond87} for arbitrary depth. For moderate viscosity (including that considered in the present work), the real part $\omega_r$ remains close to the inviscid dispersion relation (\ref{eq:SDR}), i.e. the phase velocity $c=\omega_r/k$ is marginally reduced by viscosity. The damping rate $\gamma = 2 \omega_i$ can be approximately split in two contributions: bulk dissipation due to the internal shear stress,
\begin{equation}
        \gamma_{\rm bulk} = -4\nu_\ell k^2,
\label{eq:QIT} 
\end{equation}
 and dissipation at the bottom wall due to the shear stress in the Stokes boundary layer of thickness $\delta = \sqrt{\nu_\ell /\omega_r}$ (provided that $\delta$ is much smaller than $h$)~\cite{Lighthill},
\begin{equation}
        \gamma_{\rm BW} = - \frac{ \sqrt{2 \nu_\ell \omega_r} \, k}{\sinh{2kh}}.
\label{eq:QITH} 
\end{equation}
The bulk dissipation is dominant at small wavelength, while the bottom-wall dissipation is dominant for wavelength much larger than liquid depth.

Finally, for waves in a channel of finite width $W$, a third contribution arises from the friction with the side walls~\cite{hunt1952viscous, tsai2005spatial, peirson2013rain},
\begin{equation}
        \gamma_{\rm SW} = - \frac{ \sqrt{2 \nu_\ell \omega_r}}{W}.
\label{eq:SWD} 
\end{equation}
In the following, we use the numerically resolved LeBlond-Mainardi solution, noted $\gamma_{\rm LM}$, which has the bulk and bottom-wall approximations (\ref{eq:QIT}) and (\ref{eq:QITH}) as asymptotic solutions for large and small $k$, respectively, to which we add the side-wall contribution $\gamma_{\rm SW}$ (\ref{eq:SWD}). The relative magnitude of these contributions, for the specific geometry of our experimental setup, is discussed in the Appendix.

To summarize, the net temporal growth rate $\gamma$ accessible to experiments is the sum of the positive contribution (\ref{eq:Miles}) and a combination of negative contributions (\ref{eq:QIT}) and (\ref{eq:SWD}). In most practical configurations (including the present paper), waves are forced under permanent conditions and develop spatially, so the energy transport equation (\ref{eq:EPD}) reduces to 
$$
\frac{d}{d x} E = \gamma_s E,
$$
with $\gamma_s = \gamma / c_g$ the {\it spatial} growth rate~\cite{Gaster}. Assuming homogeneous forcing and dissipation (independent of $x$), this yields an exponential variation of the energy profile, $E(x) = E_0 e^{\gamma_s x}$, with $\gamma_s > 0$ for growing waves and $\gamma_s<0$ for damped waves. In the following, we perform measurements of $\gamma_s$ by fitting $E(x)$ in the range of exponential growth or decay for various $k$ and friction velocities $u^*$, and discuss the relevance of the various growth and damping rate contributions in the case of viscous liquids.

\section{Experiments and measurements}

\subsection{Experimental setup} \label{sec:exp_setup}

\begin{figure}[tb!]
    \begin{center}
    \includegraphics[width=\textwidth]{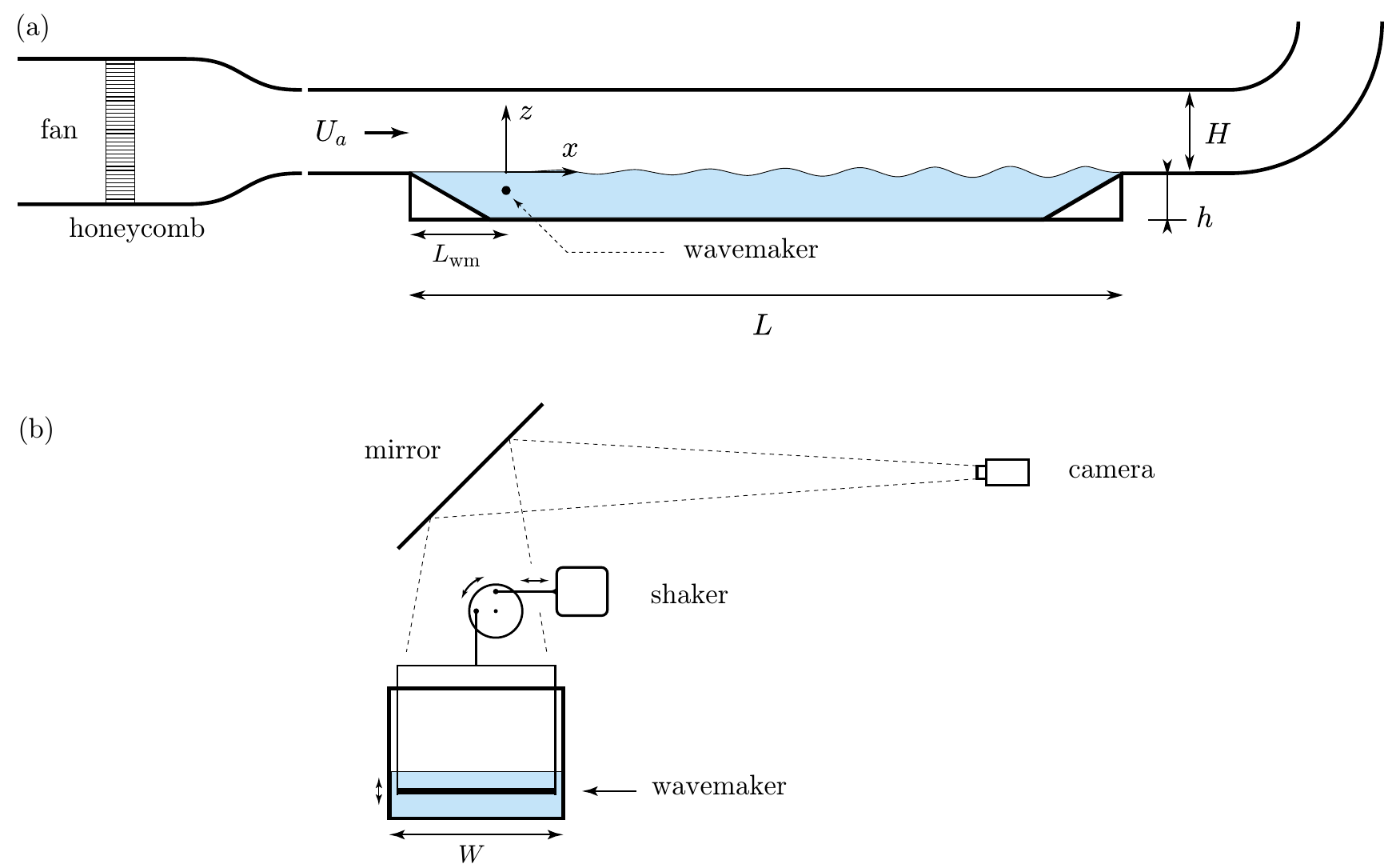}
    \caption{Sketch of the experimental setup. (a) Side view; (b) cross sectional view. $L=1500~\mathrm{mm}$, $h=35~\mathrm{mm}$, $H=105~\mathrm{mm}$, $W=296~\mathrm{mm}$. The wave maker is a cylinder of 5~mm diameter, located at $L_{wm}=125~\mathrm{mm}$ from the beginning of the tank and $-14$~mm below the liquid surface.}
    \label{fig:setup}
    \end{center}
\end{figure}

The experimental setup is composed of a rectangular tank of length $L=1.5$~m and width $W=296$~mm, located at the bottom of a wind tunnel; see~Fig.~\ref{fig:setup} and Refs.~\cite{Paquier_2015,Paquier_2016,aulnette2019wind} for details. The wind velocity $U_a$, measured in the center of the cross-section, ranges from~$0$~to~$7~\mathrm{m~s^{-1}}$. The tank, of depth $h=35~\mathrm{mm}$, is filled with silicon oil (Rhodorsil\textsuperscript{\textregistered}~oil~47~V20 and V50). Two kinematic viscosities are used: $\nu_\ell=20.5~\mathrm{mm}^2~s^{-1}$ and $\nu_\ell = 50.1\pm 0.1~\mathrm{mm}^2~s^{-1}$. The experiments are carried out at temperature of $24.0\pm1.0^\circ$C, giving an oil density $\rho_\ell=956~\mathrm{kg~m^{-3}}$ and surface tension $\sigma=
20.7~\mathrm{mN~m^{-1}}$. The corresponding capillary wavelength and capillary-gravity frequency are $\lambda_{cap}= 9.33$ mm and $f_{cap}=18.3$ Hz, and the minimum phase velocity is $c_{min} = (4 g \gamma / \rho)^{1/4} = 0.171$~m/s.

Waves are generated using an immersed wave maker, made of a vertically oscillating stainless steel cylinder of diameter $5~\mathrm{mm}$ and length matching the width of the tank. The wave maker is located $125$~mm after the beginning of the tank and defines the origin $x=0$ of our measurements. It is hung by a rectangular frame and oscillated using an electromagnetic shaker (Sinocera~JZK-20) via a crank rod system [Fig.~\ref{fig:setup}(b)]. The upper part of the cylinder is at $z(t) = z_0 + A \cos(\omega t)$, with $z_0 = - 14$~mm the mean depth and $A$ the stroke amplitude. In order to minimize wave reflection, two inclined perforated planes of length $85~\mathrm{mm}$ and slope $30^\circ$ are added at both ends of the tank to absorb the wave energy. 

The forcing frequency $f=\omega/2\pi$ ranges from $3$ to $12~\mathrm{Hz}$. This range is below the capillary-gravity crossover frequency $f_{cap}$, indicating that excited waves are in the gravity regime. The corresponding wavelengths range from 150~mm to 15~mm. The frequency range is limited at low frequency by the fact that the small diameter of the cylinder cannot excite efficiently large wavelengths, and at large frequency by the fact that the wavelength and the attenuation length become too small to be measurable. The Reynolds numbers $Re = \lambda_{cap} c_{min} / \nu_\ell$, which characterize gravito-capillary waves of wavelength $\lambda_{cap}$ propagating at the minimum phase velocity $c_{min}$, are $Re = 78$ and 32 for $\nu_\ell=20.5~\mathrm{mm}^2~s^{-1}$ and $\nu_\ell = 50.1~\mathrm{mm}^2~s^{-1}$, respectively. Since the wavelengths considered in the following are significantly larger than $\lambda_{cap}$, these values represent lower bounds for the actual Reynolds numbers, confirming that the waves are in the weakly damped regime~\cite{Leblond87}. 

The stroke amplitude of the wave maker is kept fixed to $A=1~\mathrm{mm}$ in all this study. This forcing induces waves of initial amplitude $\zeta_0$ between 0.1 and 0.4~mm depending on the forcing frequency. This initial amplitude is small enough to satisfy the linear wave approximation (the maximum wave slope near the wave maker remains below $k\zeta_0 \simeq 0.02$). On the other hand, this initial amplitude is large enough to be directly in the Miles' exponential growth regime: the characteristic surface deformation corresponding to the transition from the Phillips' incoherent wrinkles to the Miles' regular waves is estimated to $\zeta \simeq (0.11-0.14) \delta_\nu$ in Refs.~\cite{Perrard2019,li_shen_2022}, with $\delta_\nu = \nu_a / u^*$ the thickness of the viscous sublayer, which lies in the range $40-100~\mu$m.

\subsection{Friction velocity}

In order to check the scaling of the growth rate (\ref{eq:Miles}), it is necessary to determine accurately the friction velocity $u^*$ for a given wind velocity $U_a$. We deduce $u^*$ from the stationary current $U_s$ that develops at the liquid surface because of stress continuity $\rho_a u^{*2} = \rho_\ell \nu_\ell \partial u_x / \partial z |_{z=0}$. The flow in the liquid being laminar (the maximum Reynolds number $U_s h/\nu_\ell$ is less than 70), a stationary Couette-Poiseuille flow of zero flow rate develops, with a surface current $U_s$ given by~\cite{Paquier_2015}
\begin{equation}
    U_s = \frac{\rho_a u^{*2}h}{4\rho_\ell \nu_\ell}.
\label{eq:surf_vel_current}
\end{equation}
Close to the onset of wave growth, this surface current is of the order of $0.02~\mathrm{m~s^{-1}}$, which is less than 10\% of the typical phase velocity. In Aulnette {\it et al.}~\cite{aulnette2019wind}, we show that the friction velocity $u^*$ deduced from $U_s$ at various $x$ locations using Eq.~(\ref{eq:surf_vel_current}) is in good agreement with the empirical relation for a developing turbulent boundary layer -- see, e.g., Eq.~(21.12) in Schlichting~\cite{Schlichting}
\begin{equation}
      \frac{u^{*2}(x)}{{U_a}^2}\simeq C\left(\frac{(x+x_0)U_a }{\nu_a}\right)^{-0.2},
\label{eq:Sch} 
\end{equation}
with $x_0= 475$~mm the distance between the beginning of the flat plate at the end of the wind-tunnel convergent and the wave maker, and $C=0.029$.  This decrease of $u^*$ along $x$ is related to the development of the boundary layer along the tank, with a thickness that increases as $\delta(x) \simeq x^{0.8}$~\cite{aulnette2022kelvin}. In the following, we use Eq.~(\ref{eq:Sch}) to compute the friction velocity $u^*$ from the prescribed wind velocity $U_a$ close to the wave-maker, in the range $x \simeq 100-300$~mm where the growth rate is measured. The uncertainty on $u^*$ is of the order of 5\%~\cite{aulnette2019wind}.

\subsection{Wave measurement}

The surface deformations are determined using Free-Surface Synthetic Schlieren (FS-SS)~\cite{Moisy09}. This method is based on the analysis of the refracted image of a random-dot pattern through the interface. The dots are $1~\mathrm{mm}$ in diameter (corresponding to 2.3 pixels) with a density of $60$ dots per $\mathrm{cm^2}$. The pattern is printed on a transparent film located at the bottom of the liquid tank and illuminated from below by a LED panel. Images of the pattern are taken through the transparent upper-wall of the wind tunnel. To reduce parallax errors, images are acquired through a 45$^\mathrm{o}$ mirror located above the wave tank by a camera located $3~\mathrm{m}$ from the liquid surface [Fig.~\ref{fig:setup}(b)]. Images are acquired using a high-speed camera (Photron~FASTCAM~Mini~WX50), fitted with a Canon $85~\mathrm{mm}~f/1.8$ macro lens. Images of dimension of $890 \times 320~\mathrm{mm^2}$ are recorded at a frame rate at least $10$ times larger than the wave maker frequency $f$, for a recording duration of at least 20 wave periods.

The surface slope field is obtained by computing the apparent displacement field $\delta {\bf r}$ of the dot pattern induced by the surface deformation using an image correlation algorithm~\cite{Moisy09}. In the limit of weak slopes and within the paraxial approximation, the displacement field is proportional to the surface height gradient,
$$
\delta {\bf r} = - h^* \nabla\zeta,
$$
with $h^* = 14$~mm an effective distance that includes the surface-pattern distance ($h_p= 48.2~\mathrm{mm}$) and the refraction indices of the liquid and intermediate layers between the surface and the pattern~\cite{Moisy09}. Although the surface height field $\zeta({\bf r},t)$ can be reconstructed by inverting the $\nabla$ operator,  as done in Refs.~\cite{Paquier_2015,Paquier_2016}, here we work directly with the refraction-induced displacement field $\delta {\bf r}$: for a quasi-monochromatic wave propagating in the $x$ direction, $\zeta(x,t) = \zeta_0(x) \cos(kx-\omega t)$, where $\zeta_0(x)$ is a slowly varying amplitude, the displacement is simply proportional to the wave height,
$$
\delta x = - h^* \frac{\partial \zeta}{\partial x} \simeq h^* k \zeta_0(x) \sin(kx-\omega t),
$$
so the spatial growth rate of the wave amplitude $\zeta_0(x)$ can be obtained directly from that of the displacement $\delta x$.

The image correlation algorithm is based on interrogation windows of size 12 pixels (5.2~mm) with 50\% overlap, yielding a spatial resolution of $2.6~\mathrm{mm}$. This is approximately 6 times smaller than the smallest wavelength considered here, $\lambda \simeq 15~\mathrm{mm}$ for $f=12~\mathrm{Hz}$. The resolution in the apparent displacement $\delta x$ is 0.1~pixel (0.04~mm), yielding a resolution in surface slope $k\zeta \simeq 0.003$.

\subsection{Data processing} \label{sec:data_proc}

\begin{figure}[tb!]
	\begin{center}
\includegraphics[width=0.85\textwidth]{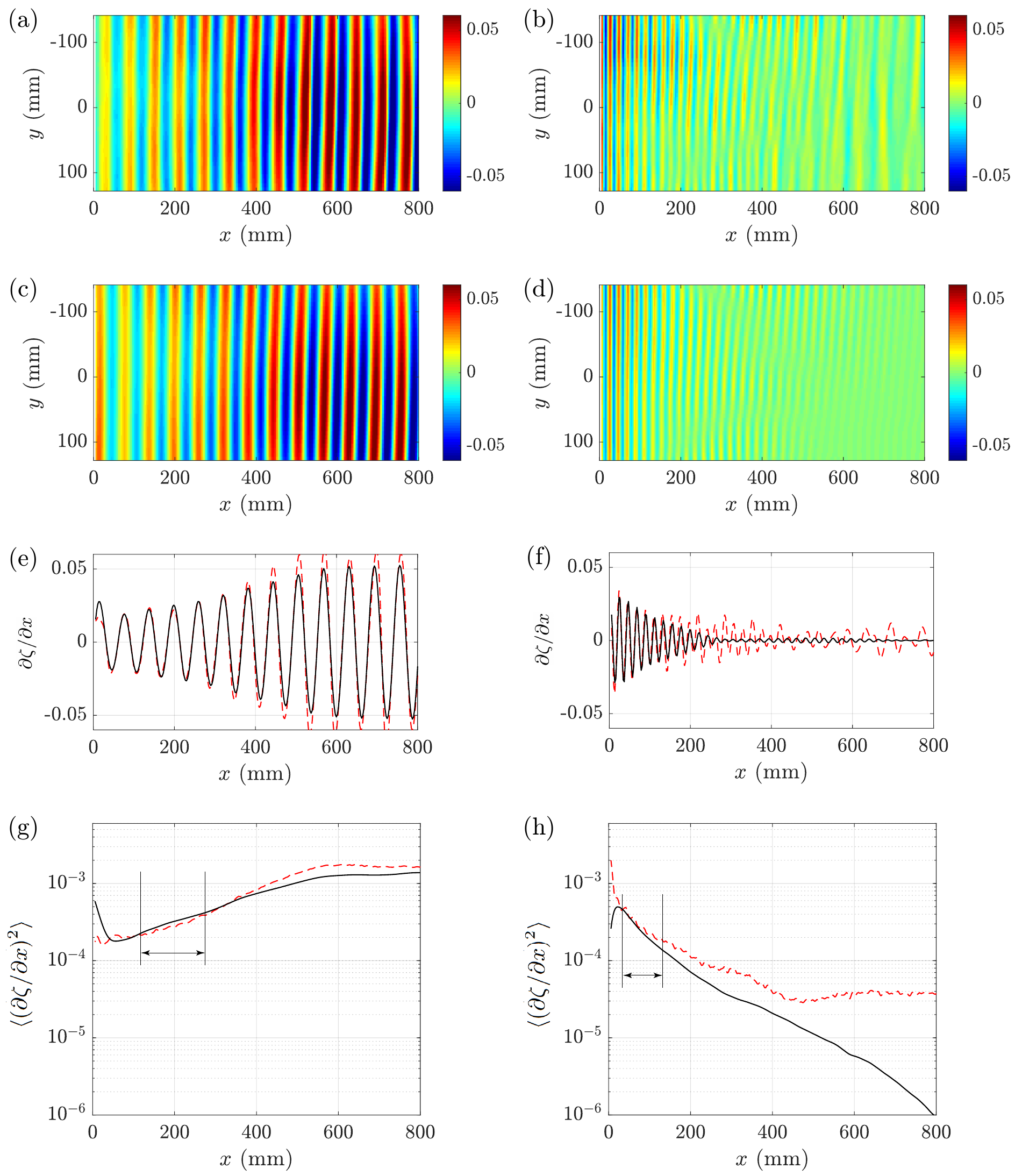}
\caption{Spatial evolution of the instantaneous and mean-square wave slope at wind velocity $U_a=5~\mathrm{m~s^{-1}}$ and oil viscosity $\nu_\ell=20.5~\mathrm{mm^2~s^{-1}}$. Left column: $f=5~\mathrm{Hz}$, for which the wave is amplified; right column: $f=9.5~\mathrm{Hz}$, for which the wave is damped. (a,b) Snapshots of the raw surface gradient $|\nabla \zeta|$. (c,d) Snapshots of the filtered surface gradient $|\nabla \bar \zeta|$ using the spatiotemporal filter (\ref{eq:Butterworth}). (e,f) Wave slope profiles $\partial \zeta / \partial x$ (red dashed curves: raw data; black continuous curves: filtered data). (g,h) Mean square envelopes $\langle (\partial \zeta / \partial x)^2 \rangle$ (raw and filtered). The  two vertical lines show the range used for the exponential fit (\ref{eq:RMS_exp}).}
\label{fig:exampleRMSField}
\end{center}
\end{figure}

Figures~\ref{fig:exampleRMSField}(a) and (b) show two snapshots of the magnitude of the wave gradient field $\nabla \zeta = - \delta {\bf r} / h^*$ measured at a wind velocity $U_a = 5.0$~m~s$^{-1}$ in the case of an amplified wave ($f=5$~Hz, left) and damped wave ($f=9.5$~Hz, right), for the oil of viscosity $\nu_\ell=20.5~\mathrm{mm^2~s^{-1}}$.  The slight tilt observed in the wave crests originates from the slightly nonhomogeneous wind in the $y$ direction, of the order of 3\%~\cite{Aulnette_PhD}. From the wave slope in the $x$ direction, we compute the mean-square envelope, which writes for a quasi-monochromatic wave as
\begin{equation}
\langle (\partial \zeta / \partial x)^2 \rangle \simeq \frac{1}{2} k^2 \zeta_0^2(x),
\label{eq:RMS_zeta}
\end{equation}
with $\zeta_0(x)$ the slowly varying wave amplitude and $\left\langle \cdot \right\rangle$ the average over time and $y$. Examples of wave slope profiles $\partial \zeta / \partial x$ at fixed $y$ and their corresponding mean-square envelopes are shown as red dashed lines in Figs.~\ref{fig:exampleRMSField}(e)-~\ref{fig:exampleRMSField}(h). In the amplified case, the envelope shows a well defined exponential growth at moderate fetch, followed by a saturation at large fetch [Fig.~\ref{fig:exampleRMSField}(g)]. This saturation, which results from a combination of nonlinear interactions with other waves and the small decrease of the wind shear stress along the developing boundary layer, is not considered in the following. The damped case is more complex [Fig.~\ref{fig:exampleRMSField}(h)]: after a short range of exponential decay, the envelope saturates or even slightly increases at large fetch, because of the presence of other waves of frequencies different from the forcing frequency. These unforced waves are the amplified waves naturally present for a wind velocity above the critical threshold, and may hide the underlying decay of the forced wave, which makes it difficult to accurately measure the damping rate.

\begin{figure}[tb!]
	\begin{center}
\includegraphics[width=0.9\textwidth]{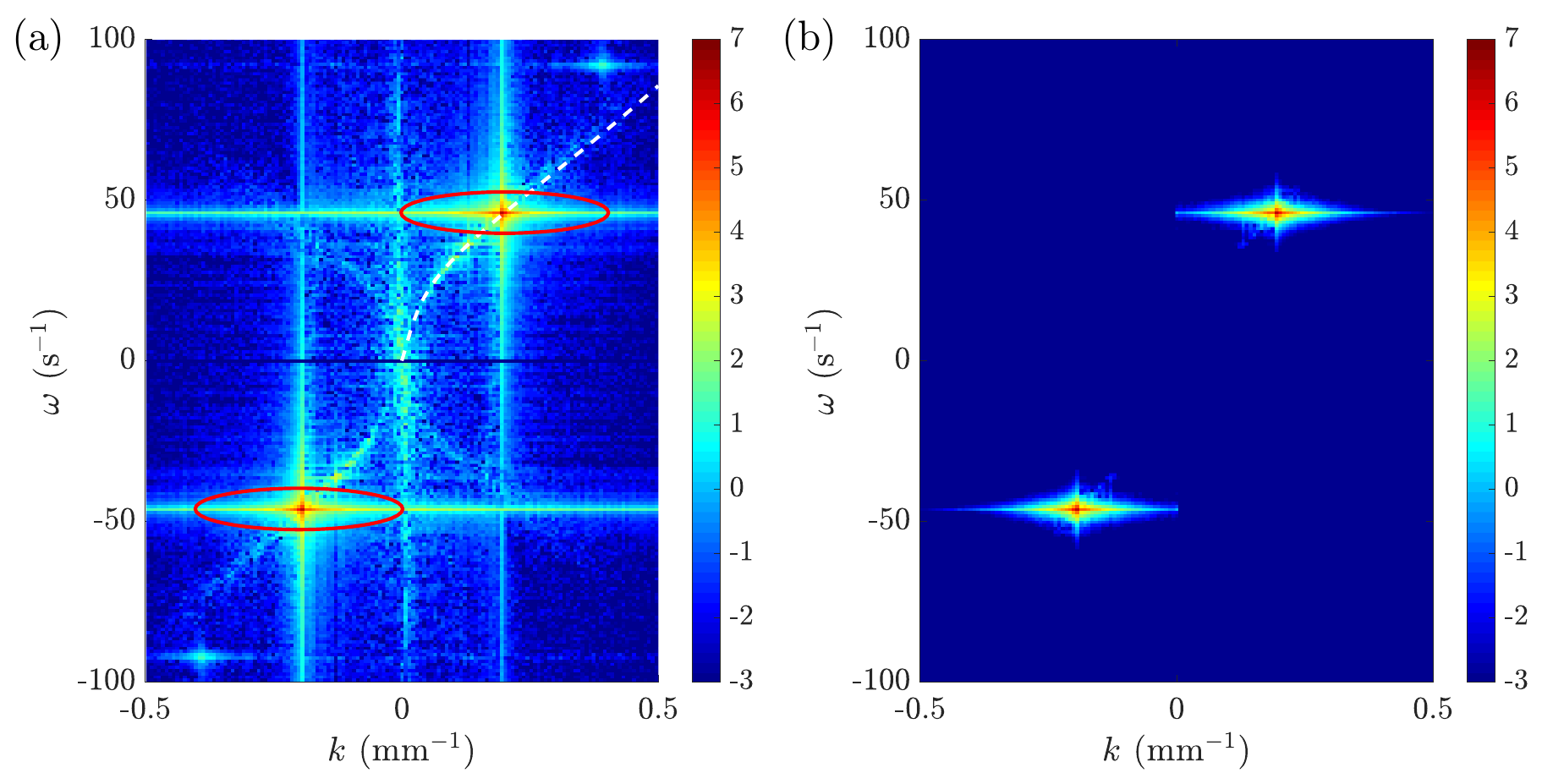}
\caption{(a) Power spectrum $|\delta \hat x|^2(k,\omega)$ of the displacement field in logarithmic scale for $U_a=4.2$~m~s$^{-1}$, $f=7.3~\mathrm{Hz}$ and oil viscosity $\nu_\ell=20.5~\mathrm{mm^2~s^{-1}}$. The white dashed line shows the viscous dispersion relation~\cite{Leblond87}. The energy of the forced wave $\pm(k_0,\omega_0)$ is selected by the spatiotemporal filter (\ref{eq:Butterworth}) (the red ellipses show the isovalue $G=0.05$). (b) Resulting filtered spectrum $|G(k,\omega) \delta \hat x|^2(k,\omega)$.}
    \label{fig:BW}
    \end{center}
\end{figure}

To extract the forced waves from the sea of ambient waves present in the field, we perform a spatiotemporal band-pass filtering of the displacement field $\delta x = -h^* \partial \zeta / \partial x$. Figure~\ref{fig:BW}(a) shows the power spectrum $|\delta \hat x|^2(k,\omega)$ (averaged in the $y$ direction) in the case of a damped wave. The displacement field being real, its Fourier transform is symmetric under $(k,\omega) \rightarrow (-k,-\omega)$. The quadrants ($+,+$) and ($-,-$) correspond to waves propagating in the $x>0$ direction, while the quadrants ($+,-$) and ($-,+$) correspond to waves propagating in the $x<0$ direction. In addition to the injected energy around the forcing $\pm (k_0, \omega_0)$ (red ellipses), energy is also visible in the reflected waves $\pm (-k_0, \omega_0)$, in harmonics, and along the viscous dispersion relation (white dashed lines). Energy along the dispersion relation is systematically found for a wind velocity above the onset, and is due to the natural waves growing from initial disturbances and amplified by the wind.

To filter out the undesired wave components, we compute for each $y$ the space-time Fourier transform $\delta \hat x(k,\omega)$, convolute it with a Butterworth filter kernel
\begin{equation}
        G\left(k,\omega\right)=\frac{H(k\omega)}{1+\left[\left(\frac{k-k_0}{\Delta k}\right)^2 + \left(\frac{\omega-\omega_0}{\Delta\omega}\right)^2\right]^{N/2}},
\label{eq:Butterworth}
\end{equation}
and compute the inverse Fourier transform to reconstruct the filtered displacement field $\delta \bar x(x,t)$, and hence the filtered wave slope $\partial \bar \zeta / \partial x$. We choose a filter kernel centered around the forcing $\pm (k_0, \omega_0)$, with a spectral bandwidth ($\Delta k$, $\Delta \omega$) and a decay characterized by the exponent $N$, chosen here equal to 6. A Heaviside operator $H(k\omega)$ is included to remove the reflected waves ($k\omega<0$). A narrow temporal bandwidth $\Delta \omega$ is chosen to select precisely the forced waves, but a wider spatial bandwidth $\Delta k$ is chosen to correctly measure the spatial decay of the wave. The temporal bandwidth is chosen as $\pm 3$ spectral points $\pi/T_{acq}$ (with $T_{acq}$ the acquisition duration) surrounding the forcing frequency $\omega_0$. The spatial bandwidth $\Delta k$ is chosen equal to $k_0$, so the smallest measurable damping length is of the order of the wavelength. The resulting filtered spectrum $| G(k,\omega) \delta \hat x|^2(k,\omega)$ in Fig.~\ref{fig:BW}(b) shows that the energy of the forced wave is correctly selected by these parameters.

The snapshots for the filtered instantaneous wave gradient $\nabla \bar \zeta = - \delta \bar {\bf r} / h^*$, shown in Figs.~\ref{fig:exampleRMSField}(c)-~\ref{fig:exampleRMSField}(d), are very close to the raw data in the amplified case, but are significantly modified in the damped case, indicating that the naturally amplified waves are efficiently filtered out by our procedure. The corresponding wave slope profiles and envelopes are also shown as continuous black lines in Figs.~\ref{fig:exampleRMSField}(e)-\ref{fig:exampleRMSField}(h).

\begin{figure}[tb!]
	\begin{center}
\includegraphics[width=0.6\textwidth]{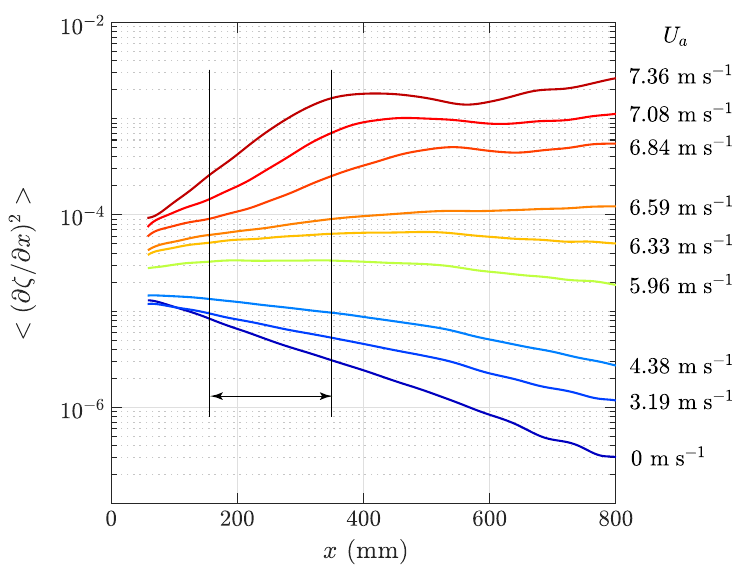}
\caption{Mean square envelopes $\langle (\partial \bar \zeta / \partial x)^2 \rangle$ (filtered data) at various wind velocities $U_a$, for $\nu_\ell = 50$~mm$^2$~s$^{-1}$ and $f=4.1$~Hz ($k = 69$~mm$^{-1}$). The vertical lines show the range used to fit the spatial growth rate $\gamma_s$ using Eq.~(\ref{eq:RMS_exp}).}
\label{fig:fitUa}
\end{center}
\end{figure}

To validate our procedure over the full range of wind velocity, we show in Fig.~\ref{fig:fitUa} a series of mean-square filtered wave slopes at a fixed frequency ($f=4.1$~Hz) for increasing $U_a$, illustrating the gradual transition from damped to amplified waves. Close to the wave maker, the growth or decay is well described by an exponential function, from which we fit the spatial growth rate $\gamma_{s}$,
\begin{equation}
    \langle (\partial \bar \zeta / \partial x)^2 \rangle \propto e^{\gamma_{s}x}.
\label{eq:RMS_exp} 
\end{equation}
We choose the range $x \in [1.8 ,4] \lambda$ for the fits [shown by the two vertical lines in Figs.~\ref{fig:exampleRMSField}(g) and \ref{fig:exampleRMSField}(h) and \ref{fig:fitUa}], for which a robust exponential fit can be performed at all $f$ and $U_a$. The overall uncertainty on $\gamma_s$ is less than 8\%.

\section{Results}
\label{secResultsDisccution}

\subsection{Phase velocity}

To compare the spatial growth rates $\gamma_s$ determined experimentally to the temporal growth rates $\gamma$ predicted theoretically, it is necessary to convert the measured spatial variations of the wave energy to the corresponding temporal variations. In the limit of small viscosity, the ratio between the two growth rates is given by the group velocity~\cite{Gaster}, which cannot be directly measured in our system. We must therefore use the group velocity obtained from the dispersion relation of the free waves in a viscous liquid, i.e. including viscous effects but without wind, as discussed in Sec.~\ref{Sec_growth_attenuation}. To check to what extent we can rely on this viscous free-wave dispersion relation, we first measure here the phase velocity of the waves and compare it with this prediction.

\begin{figure}[tb!]
\begin{center}
\includegraphics[height=0.95\textwidth]{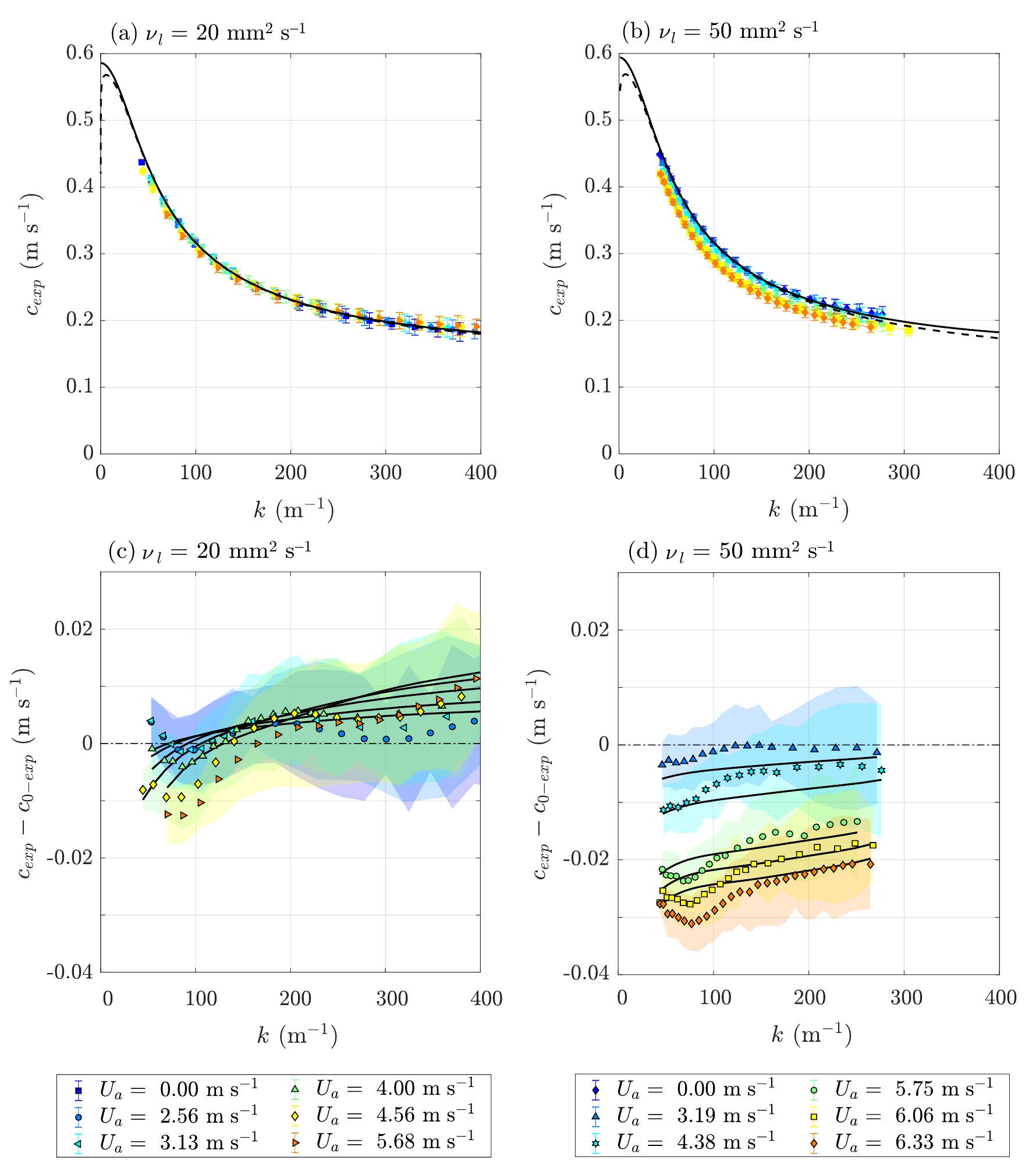}
\caption{Phase velocity as a function of the wave number for various wind velocities and for the two liquid viscosities, (a) $\nu_\ell=20.5~\mathrm{mm^2~s^{-1}}$ and (b) $\nu_\ell=50.1~\mathrm{mm^2~s^{-1}}$. Error bars correspond to the standard deviation of the data. The black lines show the phase velocity for free waves, either inviscid (solid line) or with viscosity (dashed line). (c,d) Difference between the phase velocity with and without wind. Solid lines are the predictions including the (positive) surface drift correction (\ref{eq:celerity_current}) and the (negative) aerodynamic pressure correction (\ref{eq:celerity_correction}), with uncertainties shown as color shaded areas.}
\label{fig:c_c}
\end{center}
\end{figure}

Figures~\ref{fig:c_c}(a) and \ref{fig:c_c}(b) show  the phase velocity $c_{exp} = \omega /k$ for the two liquid viscosities as a function of the wave number for various wind velocities $U_a$ (here $k$ is simply measured from the distance between wave crests, ensuring high accuracy). The data are compared to the inviscid prediction (\ref{eq:SDR}) in solid line, and to the numerically computed viscous prediction of LeBlond and Mainardi~\cite{Leblond87} including depth effects in dashed line. These predictions are very close in our range of frequencies ($3-12$~Hz), within 2\%, so we will simply consider the inviscid prediction in the following. At low frequency, the inviscid phase velocity tends to the nondispersive shallow water limit $\sqrt{gh} \simeq 0.58$~m/s, while the viscous prediction falls to 0 when the thickness of the Stokes boundary layer becomes of the order of the liquid depth (this cutoff is well below the frequencies relevant to our study).

For our range of frequencies, the measured phase velocity without wind, $c_{0-exp}$, is almost indistinguishable from the prediction (either inviscid or viscous). In the presence of wind, it still remains close to the prediction, to within 3\% at $\nu_\ell=20.5~\mathrm{mm^2~s^{-1}}$ and 8\% at $\nu_\ell=50.1~\mathrm{mm^2~s^{-1}}$. We can therefore use with reasonable accuracy the predicted group velocity computed from the inviscid dispersion relation (\ref{eq:SDR}) to convert spatial growth rates to temporal ones in the following.

Before further proceeding, it is interesting to analyze in more detail the small differences between the measured phase velocity with and without wind ($c_{exp}$ and $c_{0-exp}$, respectively). The difference $c_{exp} - c_{0-exp}$ is plotted in Fig.~\ref{fig:c_c}(c) and \ref{fig:c_c}(d), with uncertainties shown as shaded areas. It can be explained by a combination of two antagonistic effects~\cite{valenzuela1976growth}, the importance of which depending on the liquid viscosity. The first one corresponds to an increase of the phase velocity by the surface current induced by the wind shear stress, which dominates at small viscosity, and the second one corresponds to a decrease of the phase velocity due to the wave-induced aerodynamic pressure, which dominates at large viscosity.

The increase in phase velocity by surface current is simply given by $U_s$ for short wavelengths [see Eq.~(\ref{eq:surf_vel_current}), with $U_s \simeq 0.02$~m~s$^{-1}$ close to the wave onset], but is less pronounced for long wavelengths because the perturbation they induce penetrates deeper into the liquid, where the current is lower. This transport effect therefore depends on the mean velocity profile in the liquid. For the Couette-Poiseuille profile considered here~\cite{Paquier_2015}, the modification in the phase velocity has been computed by Lilly, and can be found in the appendix of the paper by Hidy and Plate~\cite{hidy1966wind},
\begin{equation}
\Delta c_w(k) = U_s\left(1-\frac{1+2\cosh{2kh}}{kh\sinh{2kh}} + \frac{3}{2k^2h^2} \right).
\label{eq:celerity_current}
\end{equation}

The second effect is the slowdown of the phase velocity by the variations of the aerodynamic pressure over the wave profile, and can be explained qualitatively as follows. The phase velocity of gravity waves is governed by gravity, i.e. by the weight of the liquid deformation. In the presence of wind, a fraction of this weight is carried by the wave-induced aerodynamic pressure (Bernoulli suction over the wave crests and over-pressure on the wave troughs), of the order of $\rho_a U_a^2 k \zeta$. The surface deformation therefore experiences an effective reduced gravity, yielding a reduced phase velocity, which can be written as~\cite{plant1980phase}
\begin{equation}
c(k) =\sqrt{c_0(k)^2-\frac{\rho_a}{\rho_\ell}U(z_k)^2},
\label{eq:celerity_correction}
\end{equation}
with $U(z_k)$ the wind velocity evaluated at a characteristic elevation that depends on the wave number $k$. We evaluate this effect by considering a classical logarithmic profile $U(z)=u^*[\kappa^{-1}\log(zu^*/\nu_a)+B]$ (with $\kappa= 0.41$ and $B=3.5$) computed at the elevation $z_k = \epsilon/k$ with $\epsilon \simeq 0.23$~\cite{aulnette2022kelvin}.

The correction to the phase velocity, combining the surface current effect (\ref{eq:celerity_current}) and the aerodynamic pressure effect (\ref{eq:celerity_correction}), is plotted in Figs.~\ref{fig:c_c}(c) and \ref{fig:c_c}(d). It provides a reasonable description of the data: At low viscosity and large wave numbers, the waves are essentially transported by the surface current, yielding a phase velocity larger than the inviscid prediction. On the other hand, waves at low wave numbers are systematically slower than the free-wave celerity, as a result of the dominant aerodynamic pressure effect.

To summarize, the phase velocity of the waves in presence of wind can be correctly described by the finite-depth inviscid phase velocity, by including the corrections due to the surface drift and the aerodynamic pressure. However, in the range of liquid viscosities and wind velocities considered here, these corrections remain small: for a wind velocity close to the wave onset, the measured phase velocity matches the free-wave inviscid prediction to within 10\%. By extension, we can consider that the group velocity is also only marginally affected by the surface current and aerodynamic pressure, which allows us to use the free-wave inviscid prediction $c_g$ derived from Eq.~(\ref{eq:SDR}) to infer the temporal growth rate $\gamma$ from the measured spatial growth rate $\gamma_s$.

\subsection{Spatial growth rates} 
\label{Sec_Growth_rate}

\begin{figure}[tb!]
    \begin{center}
    \includegraphics[width=0.9\textwidth]{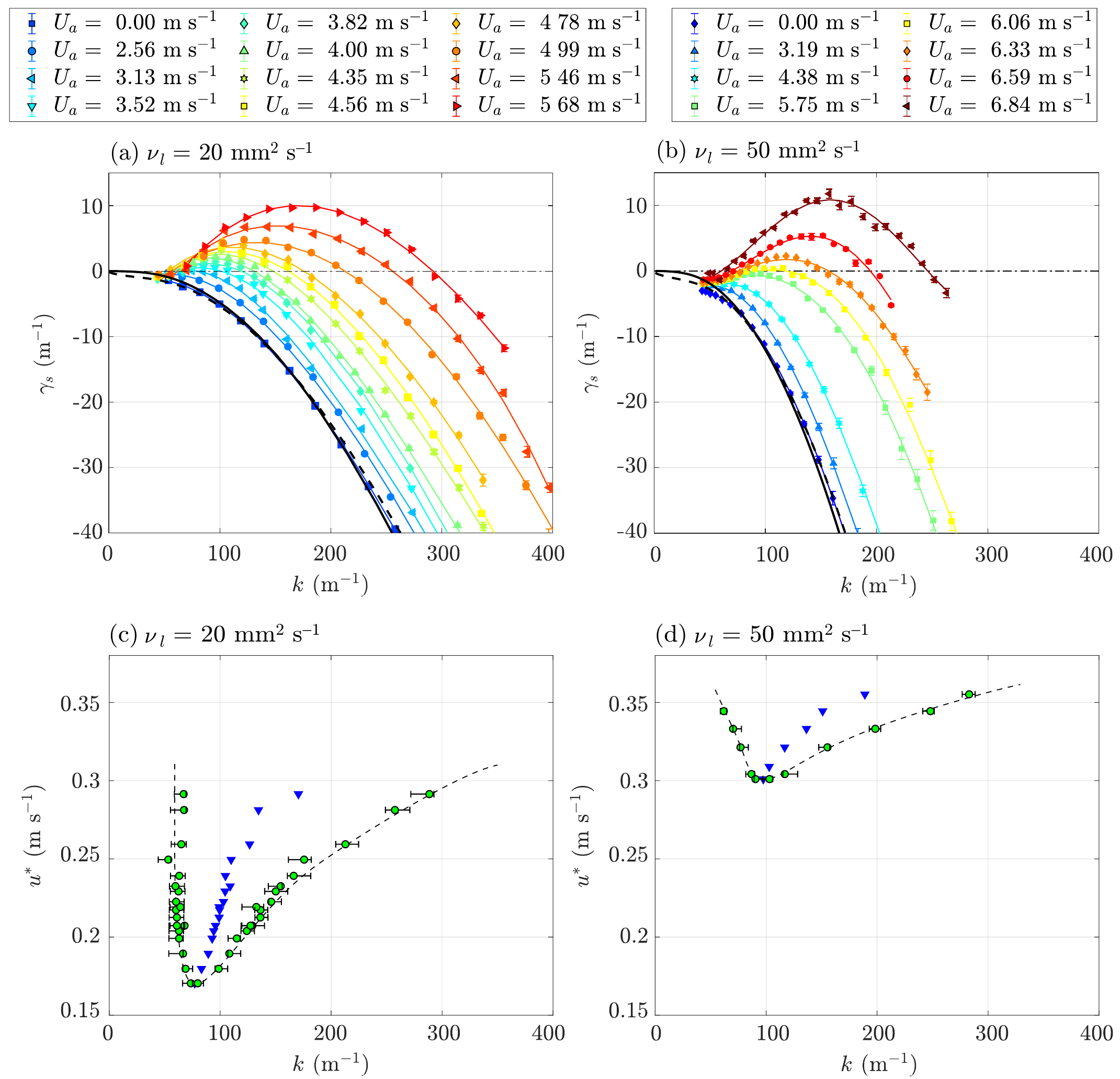}
    \caption{(a,b) Spatial growth rate $\gamma_{s}$ as a function of the wave number for various wind velocities, for (a) $\nu_\ell=20.5~\mathrm{mm^2~s^{-1}}$ and (b) $\nu_\ell=50.1~\mathrm{mm^2~s^{-1}}$. Black solid line: bulk dissipation ($-4 \nu_\ell k^2/c_g$); black dashed line: numerically solved LeBlond-Mainardi dissipation (bulk and bottom-wall dissipation) combined with side-wall dissipation.  Color lines: local polynomial regression (``loess'' smoothing). (c,d) Marginal stability curve (green circles) and most amplified wave number (blue triangles). Thin dashed lines are guides for the eyes.}
    \label{fig:beta_sm}
    \end{center}
\end{figure}

Figures~\ref{fig:beta_sm}(a) and \ref{fig:beta_sm}(b) shows the measured spatial growth rate $\gamma_{s}$ as a function of the measured wave number, for different wind velocities. As expected, without wind $\gamma_s$ is negative for all $k$. The corresponding damping rate is compared to the bulk prediction $\gamma_s = -4 \nu_\ell k^2 / c_g$ (solid line) and to the numerically solved LeBlond-Mainardi prediction~\cite{Leblond87} (including bulk and bottom-wall dissipation) completed with side-wall dissipation (\ref{eq:SWD}) (dashed line). The large wave numbers ($k>70$~m$^{-1}$) are equally well described by the two models, but the smaller $k$ are much better described by the refined model, especially at large viscosity, highlighting the significant effect of the bottom dissipation at small $k$ (see Appendix~\ref{Appendix_contributions}).

\begin{table}[t!]
\caption{Critical parameters at the onset of wave generation for the two liquid viscosities.}
\begin{center}
\begin{tabular}{p{2.5cm}p{2.5cm}p{3cm}p{2.5cm}p{2.5cm}}
    \hline
    \hline
     $\nu_\ell~(\mathrm{mm^2~s^{-1}})$ & $U_{ac}~(\mathrm{m~s^{-1}})$ & $u_c^*~(\mathrm{m~s^{-1}})$ & $k_c$ (m$^{-1}$) & $f_c$~(Hz) \\ 
     \hline
     $20.5$ & $3.18~\pm 0.04$ & $0.173~\pm0.005$ & $80 \pm 5$ & $4.35~\pm 0.1$ \\ 
     $50.1$ & $5.92~\pm 0.04$ & $0.303~\pm0.005$ & $100 \pm 6$ & $4.60~\pm 0.1$ \\ 
    \hline
    \hline
\end {tabular}
\end{center} 
\label{Table-onset}
\end{table}

As the wind velocity is increased, the waves become gradually less damped and, above a critical wind $u^*_c$, the growth rate becomes positive in a finite range of wave numbers. We recall that the friction velocity $u^*$ is deduced from the free-stream velocity $U_a$ by using Eq.~(\ref{eq:Sch}) for $x$ close to the wave maker.  From the set of growth rate curves $\gamma_s(k)$ at various velocities, we can compute the marginal stability curve $u^*(k)$, defined as the minimum friction velocity required for a given wave number $k$ to become unstable. It is plotted in Figs.~\ref{fig:beta_sm}(c) and \ref{fig:beta_sm}(d) for the two liquid viscosities.   The lowest point $(k_c, u^*_c)$ of the marginal stability curve defines the critical wave number and critical friction velocity, which are summarized in Table \ref{Table-onset} for the two liquid viscosities. The critical wave numbers correspond to wavelengths $\lambda_c \simeq 60-80$~mm, much larger than the capillary wavelength $\lambda_{cap}= 9.33$~mm.

As the liquid viscosity is increased, the critical friction velocity strongly increases, but the critical wave number is marginally modified. Increasing $u^*$ shifts the most amplified wave number (shown as blue triangles) to larger values, and rapidly widens the domain of unstable wave numbers. The low-$k$ branch of the stability curve, for which the bottom-wall and the bulk dissipation are of comparable importance (see the Appendix), shows a weak dependence with $u^*$. On the other hand, the large-$k$ branch, which is essentially governed by the bulk dissipation, strongly varies with $u^*$. We note that this higher branch is defined with high accuracy, because $\gamma_s$ sharply crosses 0, but that the lower branch is more difficult to measure, because $\gamma_s$ shows little variations with $u^*$ for small $k$.

\subsection{Comparison with Miles' scaling}
\label{Sec_Miles}

\begin{figure}[tb!]
	\begin{center}
\includegraphics[width=0.90\textwidth]{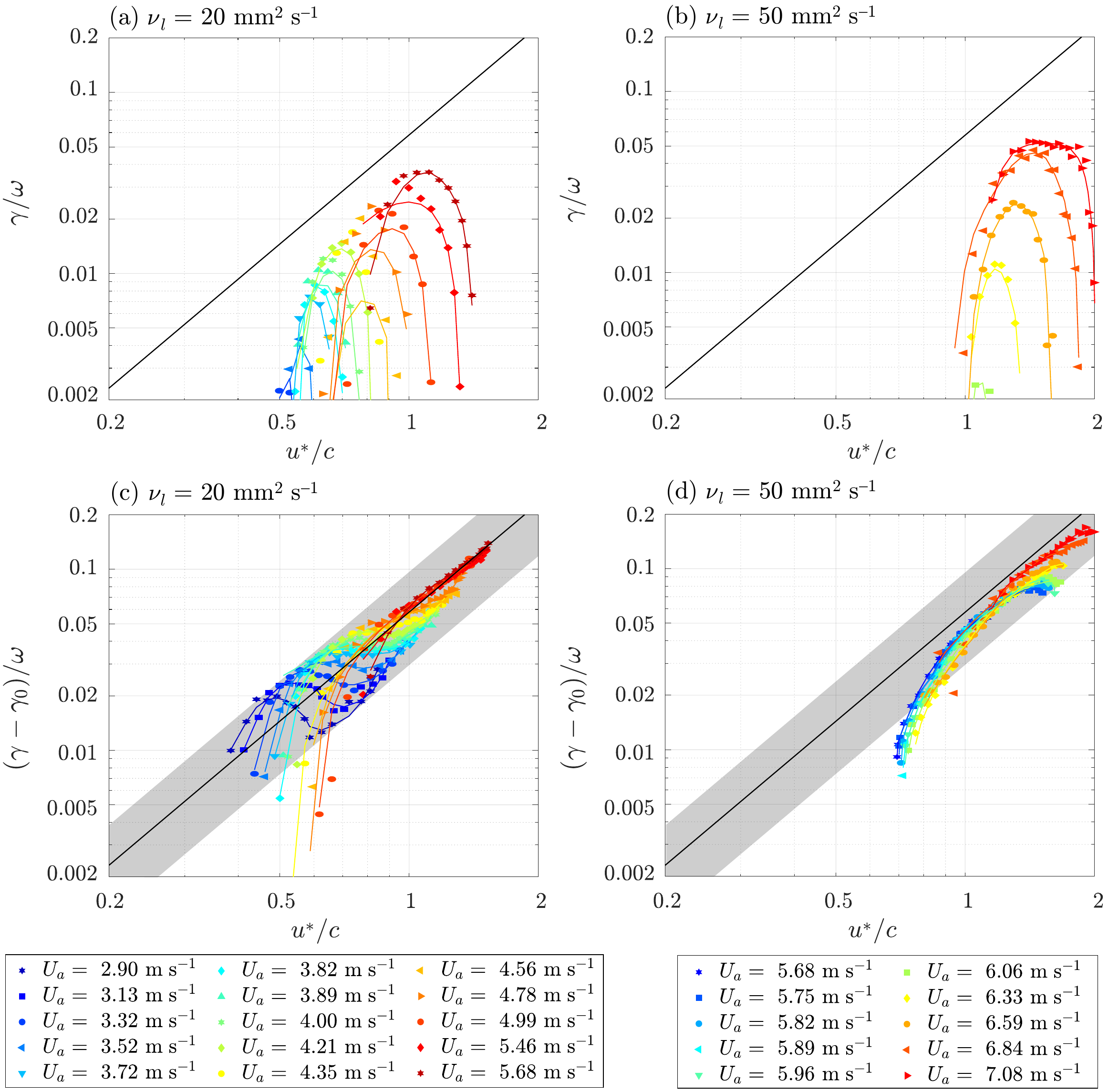}
 \caption{Dimensionless growth rate $\gamma/\omega$ as a function of the inverse wave age $u^*/c$, for (a,c) $\nu_\ell = 20.5$, and (b,d) 50.1~$\mathrm{mm^2~s^{-1}}$. The solid line shows Eq.~(\ref{eq:Miles}) for $\beta = 45$. (a, b) Raw growth rates, without viscous correction. (c, d) Growth rates corrected by the viscous contribution $\gamma_0$ measured without wind. The shaded area depicts the range $\beta = 23-75$.}
\label{fig:beta_uc}
\end{center}
\end{figure}

We now investigate to what extent Miles' model, usually applied to the air-water interface, also applies to the case of the more viscous liquids considered here. Following Eq.~(\ref{eq:Miles}), we plot in Fig.~\ref{fig:beta_uc}(a) and \ref{fig:beta_uc}(b) the temporal growth rate $\gamma= \gamma_s c_g$ normalized by the wave frequency $\omega$ as a function of the inverse wave age $u^*/c$ (only amplified waves are considered here). In this traditional representation, ``young'' waves (high $k$, low $c$) are on the right and ``old'' waves (small $k$, large $c$) are on the left. However we note that, strictly speaking, there is wave aging in our experiments: each data point corresponds to a single wave excited at a given frequency.

\begin{figure}[tb!]
\begin{center}
\includegraphics[width=0.45\textwidth]{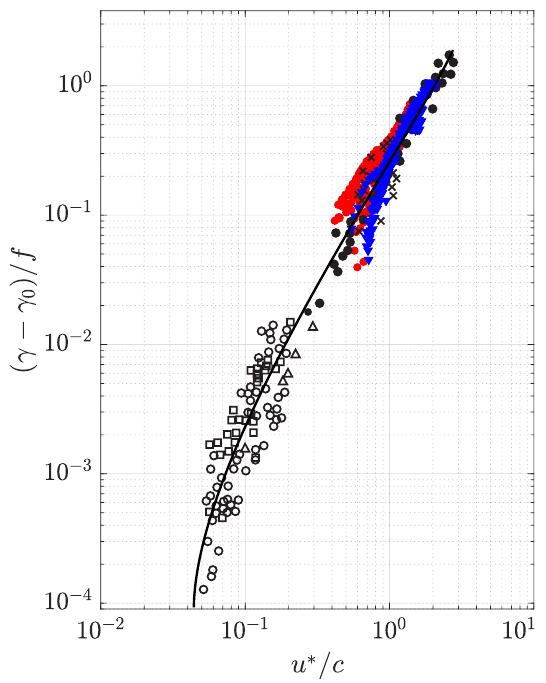}
\caption{Dimensionless net growth rate $(\gamma-\gamma_0)/f$ (with $\gamma_0$ the damping rate without wind) vs the inverse wave age, showing our data (red circles:~$\nu_\ell=20.5~\mathrm{mm^2~s^{-1}}$, blue triangles:~$\nu_\ell=50.1~\mathrm{mm^2~s^{-1}}$) superimposed to data obtained in water and compiled by Plant~\cite{plant1982relationship} (see also~\cite{komen1996dynamics, Janssen_2004, melville2015equilibrium}). Open symbols: field data; filled symbols and crosses: laboratory data; solid line: Miles's prediction~\cite{Miles1959_part2}. }
\label{fig:comparaisonMiles}
\end{center}
\end{figure}

Although individual curves for each $u^*$ fail to follow the $(u^*/c)^2$ scaling, the upper envelope of the set of curves is compatible with this scaling. The strong departure at small and large wave age results from a combination of dissipation effects (dominated by the bulk dissipation for large $u^*/c$ and bottom-wall dissipation for small $u^*/c$) and the selective amplification of the Miles' mechanism: wave numbers $k$ far from the most amplified wave numbers correspond to a critical height where the curvature of the mean velocity profile $U''(z)$, and consequently the growth rate, becomes zero.

To further test the $(u^*/c)^2$ scaling, it is therefore necessary to subtract the contribution from the viscous dissipation. In order not to rely on a specific dissipation model, we use the measured damping rate $\gamma_0$ without wind as the best estimate for the global dissipation. We therefore plot the normalized net growth rate $(\gamma - \gamma_0)/\omega$ in Figs.~\ref{fig:beta_uc}(c) and \ref{fig:beta_uc}(d), which is now consistent with the scaling $(u^*/c)^2$, at least for large wind velocity and large wave numbers (large $u^*/c$). The best fit with Eq.~(\ref{eq:Miles}) gives the average value $\bar \beta \simeq 45$ with an uncertainty range $\beta \simeq 23-75$ (shown in the shaded area).

The significant scatter in the dimensionless growth rates is not specific to our experiments performed with viscous oils, and is also present in experiments in the air-water case. In Fig.~\ref{fig:comparaisonMiles} we superimpose our normalized net growth rates to the classical Plant's compilation of laboratory and field measurements~\cite{plant1982relationship}, reproduced from Janssen~\cite{Janssen_2004}. The correct agreement between our data in viscous oils and the data in water supports the robustness of the scaling $(u^*/c)^2$ at large $u^*/c$, with a comparable scatter. The solid line shows the prediction of Miles~\cite{Miles1959_part2}, which gives the asymptotic scaling $(u^*/c)^2$ at large $u^*/c$, relevant to our ``young'' (slow) waves, but a lower growth rate for ``older'' (faster) waves when the phase velocity becomes of the order of the wind velocity.

\subsection{Maximum growth rate}

\begin{figure}[tb!]
	\begin{center}
\includegraphics[width=0.95\textwidth]{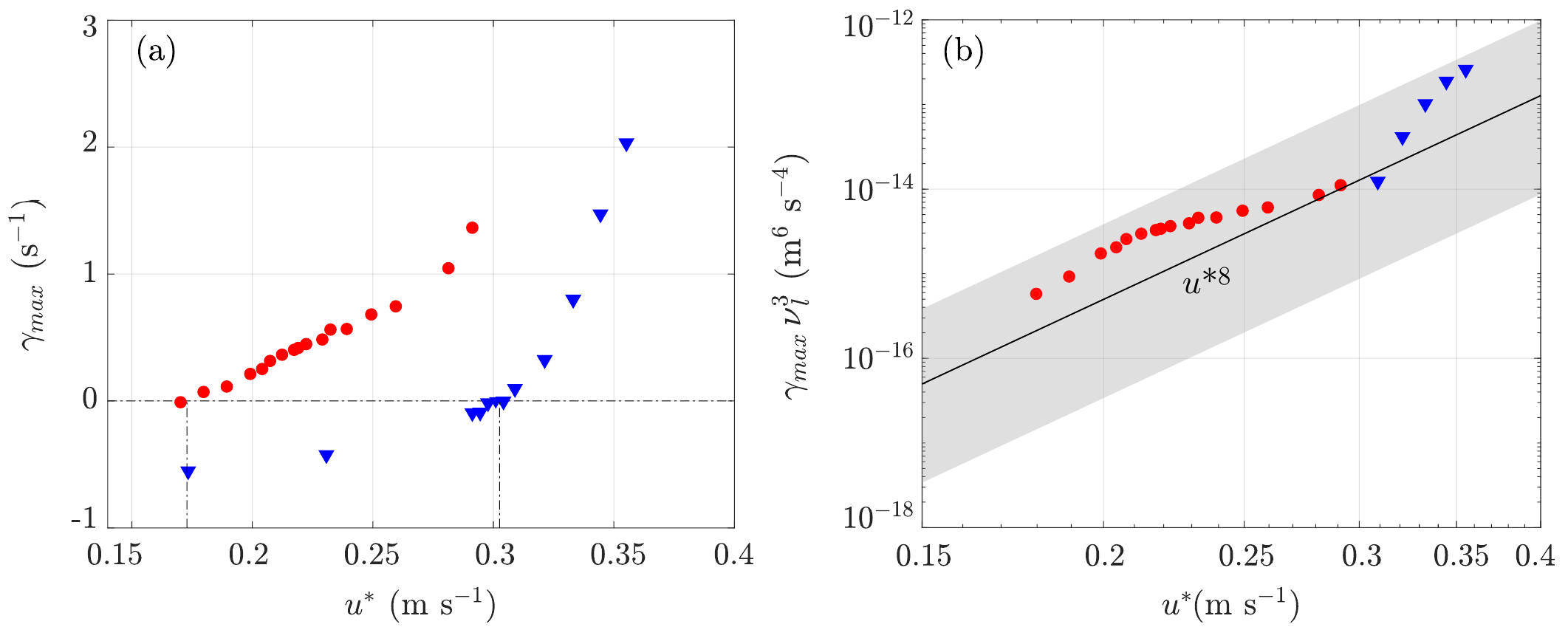}
\caption{(a) Maximum temporal growth rate $\gamma_{max}$ as a function of the friction velocity $u^*$ for $\nu_\ell=20.5~\mathrm{mm^2~s^{-1}}$ (red) and $\nu_\ell=50.1~\mathrm{mm^2~s^{-1}}$ (blue). 
Vertical dash-dotted lines show the critical friction velocities $u_c^*$~(Table~\ref{Table-onset}). (b) Viscosity-weighted maximum temporal growth rate $\gamma_{max}\nu_{\ell}^3$ as a function of the friction velocity $u^*$. The line corresponds to Eq.~(\ref{eq:beta_max_gravity}) with $\bar \beta = 45$, and the shaded area shows the range $\beta = 23-75$.}
\label{fig:beta_ut}
\end{center}
\end{figure}

We now turn to the maximum growth rate $\gamma_{max}$ (growth rate of the most amplified wave number) and its dependence on the liquid viscosity. Figure~\ref{fig:beta_ut}(a) shows $\gamma_{max}$ as a function of the friction velocity $u^*$ for the two viscosities. By definition $\gamma_{max}$ increases with $u^*$ and crosses 0 at the critical friction velocity $u^*_c$ (given in Table~\ref{Table-onset}). We note that negative values of $\gamma_{max}$ can be defined for $u^* < u^*_c$ because slightly below $u^*_c$ the growth rate shows a local (negative) maximum, corresponding to the least damped wave number.

Given the various sources of dissipation, with distinct scaling laws with respect to $\nu_\ell$ (bulk dissipation in $\nu_\ell$ and wall dissipation in $\nu_\ell^{1/2}$), a unique scaling for $\gamma_{max}$ with $\nu_\ell$ cannot be derived in general. However, an approximate trend can be proposed, at the cost of a number of approximations: (1) we ignore the $k$ dependence of the parameter $\beta$ in Eq.~(\ref{eq:Miles}), which is acceptable close to the most amplified $k$, and consider the average value $\bar \beta=45$; (2) we consider that waves are in the deep-water gravity regime, $\omega = \sqrt{gk}$, which again is acceptable because $k_c / k_{cap} < 0.15$ and $\tanh(k_c h) \simeq 0.99$; and (3) we consider the bulk dissipation $\gamma_{\rm bulk} = -4 \nu_\ell k^2$ as the only source of dissipation. This third approximation is the most questionable in our experiments, with a bulk-bottom transition at $k\simeq 70$~m$^{-1}$ (see the Appendix), which is close to the critical wave number $k_c$. Under these approximations, the temporal growth rate can be modeled as
\begin{equation}
\gamma \simeq \bar \beta s \frac{u^{*2} k^{3/2}} {g^{1/2}} - 4 \nu_\ell k^2,
\label{eq:gam_bulk}
\end{equation}
with $s = \rho_a / \rho_\ell$ the density ratio. Solving for $\partial \gamma / \partial k = 0$ yields the maximum growth rate
\begin{equation}
    \gamma_{max}=\frac{3^3}{2^{14}}  \bar \beta^4 s^4 \frac{u^{*8}}{g^2\nu_{\ell}^3}.
\label{eq:beta_max_gravity}
\end{equation}

To verify this scaling, we plot in Fig.~\ref{fig:beta_ut}(b) the combination $\gamma_{max} \nu_\ell^3$ as a function of $u^*$. The data for the two viscosities approximately gather along the power law $u^{*8}$, with variations of the same order as the uncertainty on the parameter $\beta$ found previously (the black line gives the average value $\bar \beta = 45$, and the shaded area shows the same range of $\beta$ as in Fig.~\ref{fig:beta_uc}). Although the data for the two viscosities do not overlap, because of the very limited range of accessible $u^*$ (at most twice the critical value $u^*_c$), the high-$\nu_\ell$ curve reasonably extends the low-$\nu_\ell$ curve, which is consistent with the scaling law (\ref{eq:beta_max_gravity}).

The variation of $\gamma_{max}$ with $u^*$ has been the subject of many studies, but only in the air-water case~\cite{larson1975wind,Kawai1979generation,creamer1992surface,tsai2004stability,zeisel2008viscous}. Approximate power laws $\gamma_{max} \sim u^{*n}$ are reported, but with significantly smaller exponents $n$, in the range 1.5--3.5. The asymptotic scaling $n=8$ may be difficult to access experimentally in water experiments because nonlinear effects (wave saturation, frequency downshift, wave breaking) rapidly appear as $u^*$ is increased, even close to the onset $u^*_c$.

\subsection{Critical friction velocity}

\begin{figure}[tb!]
\begin{center}
\includegraphics[width=0.50\textwidth]{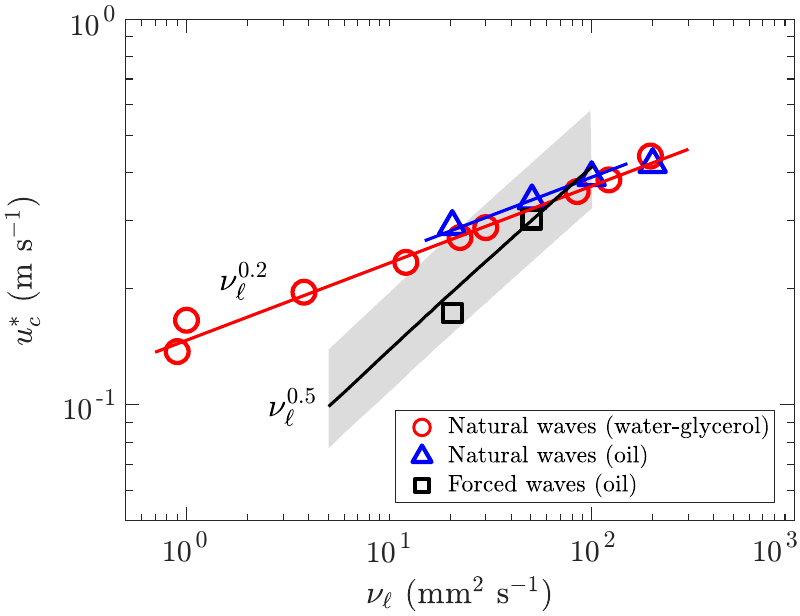}
\caption{Comparison between the critical friction velocity $u^*_c$ for natural and forced waves. $\circ$, Natural waves in water-glycerol mixtures~\cite{Paquier_2016}. $\triangle$, Natural waves in silicon oils~\cite{aulnette2022kelvin}. $\square$, Forced waves in silicon oils (present data). The black line shows the numerical solution of $\gamma=0$ using Eq.~(\ref{eq:gam_lm}) for $\bar \beta = 45$, and the shaded area shows the range $\beta = 23-75$.}
\label{fig:u_viscosite}
\end{center}
\end{figure}

We finally consider the influence of the liquid viscosity on the critical friction velocity $u^*_c$ for the growth of mechanically generated waves. This critical friction velocity cannot be directly inferred from Eq.~(\ref{eq:gam_bulk}), based on the sole bulk dissipation, because it predicts a range of unstable $k$ starting from $k=0$ as soon as $u^*>0$. This directly follows from the approximation $\beta(k) \simeq \bar \beta$ and the omission of other dissipation sources. In our experiments, the dissipation being given by a combination of bulk and bottom-wall contributions (see the Appendix), we can modify Eq.~(\ref{eq:gam_bulk}) by still assuming a constant coefficient $\beta$, while now including the Leblond-Mainardi damping rate
\begin{equation}
\gamma \simeq \bar \beta s \frac{u^{*2} k^{3/2}} {g^{1/2}} + \gamma_{\rm LM}(k).
\label{eq:gam_lm}
\end{equation}
The critical friction velocity $u_c^*$, obtained by solving numerically $\gamma=0$, is plotted in Fig.~\ref{fig:u_viscosite} as a function of $\nu_\ell$. The numerical solution, shown again for the average value $\bar \beta=45$ (black line) and its uncertainty range (shaded area), is in good agreement with the measured $u_c^*$ for the two values of $\nu_\ell$.

Interestingly, the measured $u^*_c$ and the numerical solution are well described by the approximate power law $u^*_c \sim \nu_\ell^{1/2}$. Although this power law cannot be directly derived from Eq.~(\ref{eq:gam_lm}), it is compatible with the fact that the most unstable wave number $k_c$ shows little variation with $\nu_\ell$. Considering $k_c$ constant in Eq.~(\ref{eq:gam_bulk}) and solving for $\gamma=0$ indeed yields
\begin{equation}
u^*_c \simeq \sqrt{\frac{4 \nu_\ell \omega_c}{\bar \beta s}},
\label{eq:wd}
\end{equation}
with $\omega_c = c k_c$ the frequency of the critical wavenumber. This is equivalent to the relation proposed by Wu and Deike~\cite{Wu_2021}, $u^* / c \simeq a \sqrt{\nu_l k^2 T}$, with $T=2\pi/\omega$ the wave period and $a$ a numerical factor. The best fit to our data using this relation yields $a \simeq 3.0 \pm 0.2$, which is a factor 2 below the value $a \simeq 6.76$ found in the numerical simulations of Ref.~\cite{Wu_2021}. In spite of this difference, probably related to the difference in the velocity profile (which is linear in the numerics), this suggests that the scaling (\ref{eq:wd}) for the critical friction velocity in the case of mechanically generated waves is robust.

We close this discussion by comparing this critical friction velocity and the one obtained for natural waves, without mechanical forcing. In Fig.~\ref{fig:u_viscosite} we add $u^*_c$ for natural waves as measured in the same setup in Refs.~\cite{Paquier_2016,aulnette2022kelvin}, using silicon oils and water-glycerol mixtures. Here $u^*_c$ is defined as the transition from the small-amplitude wrinkles (Phillips regime) to the exponentially growing waves (Miles regime). This natural-wave threshold is larger than the forced-wave threshold, but shows a shallower dependence with the liquid viscosity, $u^*_c \sim \nu_\ell^{0.2}$, with a very similar numerical prefactor (within 5\%) for the two types of fluids.

The two-stage scenario provides a natural framework to interpret this difference: 
Mechanically-generated waves grow from a disturbance of initial amplitude sufficient to disturb the air flow and trigger the Miles instability mechanism. On the other hand, natural waves grow from a noisy base state defined by the low-amplitude incoherent wrinkles excited by the turbulent pressure fluctuations in the air, so they require a larger wind velocity for the wrinkle amplitude to disturb the flow and trigger the instability. More specifically, for $u^* < u^*_c$, the wrinkle amplitude results from the balance between the power injected by the turbulent pressure fluctuations and the viscous dissipation in the liquid, yielding $\zeta_w \simeq s \delta (u^{*3}/g \nu_\ell)^{1/2}$, with $\delta$ the boundary layer thickness. Assuming a Phillips-Miles transition when the wrinkle amplitude $\zeta_w$ becomes of the order of the viscous sublayer thickness $\delta_\nu = \nu_a / u^*$~\cite{Perrard2019,li_shen_2022}, we obtain $u^*_c \sim \nu_\ell^{1/5}$, in good agreement with the data in Fig.~\ref{fig:u_viscosite}. The strong difference between these two scalings, $u^*_c \sim \nu_\ell^{1/2}$ for forced waves and $u^*_c \sim \nu_\ell^{1/5}$ for natural waves, emphasizes the difficulty in defining a unique onset for the wave generation problem, and the interest of varying the liquid viscosity to gain more insight into this problem.

\section{Conclusion}
\label{sec:conclusion}

In this paper we investigated the wind-induced growth of mechanically generated waves over liquids of viscosity 20 and 50 times larger than that of water, with the aim to gain insight into the scaling of the growth rate in the exponential regime. This growth rate has been extensively studied in the literature but only on the air-water configuration. Varying the liquid viscosity offers an interesting opportunity to test existing theories and shed light on this problem.

Using Free-Surface Synthetic Schlieren with a spatio-temporal filtering provides wave slope measurements with a very good resolution, and the possibility to access the growth and damping rates of the forced waves even when masked by other naturally growing waves. From these measurements we reconstruct the marginal stability curves with unprecedented accuracy. Once the viscous dissipation is subtracted, our measurements show correct overall agreement with Miles' scaling, with a significant scatter comparable to that of experiments performed in water. Our analysis suggests a maximum growth rate (growth rate of the most amplified wave number) that scales as $u^{*8} \nu_\ell^{-3}$, in reasonable agreement with our data.

Our work raises the question of the origin of the critical friction velocity $u^*_c$ for wave growth, and its dependence with liquid viscosity $\nu_\ell$. For mechanically generated waves, provided that the initial wave amplitude is not too small, the system is directly in the second stage of wave growth (exponential amplification of the wave energy). In this case, the threshold is governed by the balance between the power injected by the wind and the dissipation in the liquid. On the other hand, for natural waves, a larger threshold $u^*_c$ is found, because the wave instability is triggered from the low-amplitude incoherent wrinkles excited by the turbulent pressure fluctuations in the air~\cite{Perrard2019,li_shen_2022}. Previous works~\cite{Paquier_2016,aulnette2022kelvin} suggested a dependence $u^*_c \sim \nu_\ell^{1/5}$ for natural waves, while the present work, available for two liquid viscosities only, suggests a dependence $u^*_c \sim \nu_\ell^{1/2}$ for mechanically generated waves,  in qualitative agreement with recent numerical results~\cite{Wu_2021}. Extending these results to arbitrary liquid depth and to a larger range of viscosities, including the case of water, would be valuable to clarify this difference.

\begin{acknowledgments}

We are grateful to F. Charru, J. Magnaudet, R. Mathis, F. Burdairon, M. Aulnette and S. Perrard for fruitful discussions, and to P. Balondrade and J. Sant-Anna for assistance with measurement and data processing. We thank A. Aubertin, L. Auffray, J. Amarni and R. Pidoux for experimental help. This work was supported by the project ``ViscousWindWaves'' (ANR-18-CE30-0003) of the French National Research Agency.

\end{acknowledgments}

\appendix

\section*{Appendix: Relative importance of the various contributions to the wave dissipation}
\label{Appendix_contributions}

The damping rate of waves at the surface of a viscous liquid in a container, provided that the thickness of the boundary layers is much smaller than the container dimension, can be split into three contributions: bulk ($\gamma_{\rm bulk}$), bottom-wall ($\gamma_{\rm BW}$) and side-wall ($\gamma_{\rm SW}$), given by Eqs.~(\ref{eq:QIT}), (\ref{eq:QITH}), and (\ref{eq:SWD}), respectively. The first two contributions are contained in the finite-depth damping rate $\gamma_{\rm LM}$ for arbitrary viscosity derived by LeBlond and Mainardi~\cite{Leblond87}, which has $\gamma_{\rm bulk}$ and $\gamma_{\rm BW}$ as asymptotic solutions at low viscosity for large and small $k$ respectively.

\begin{figure}[b!]
    \begin{center}
        \includegraphics[width=0.92\textwidth]{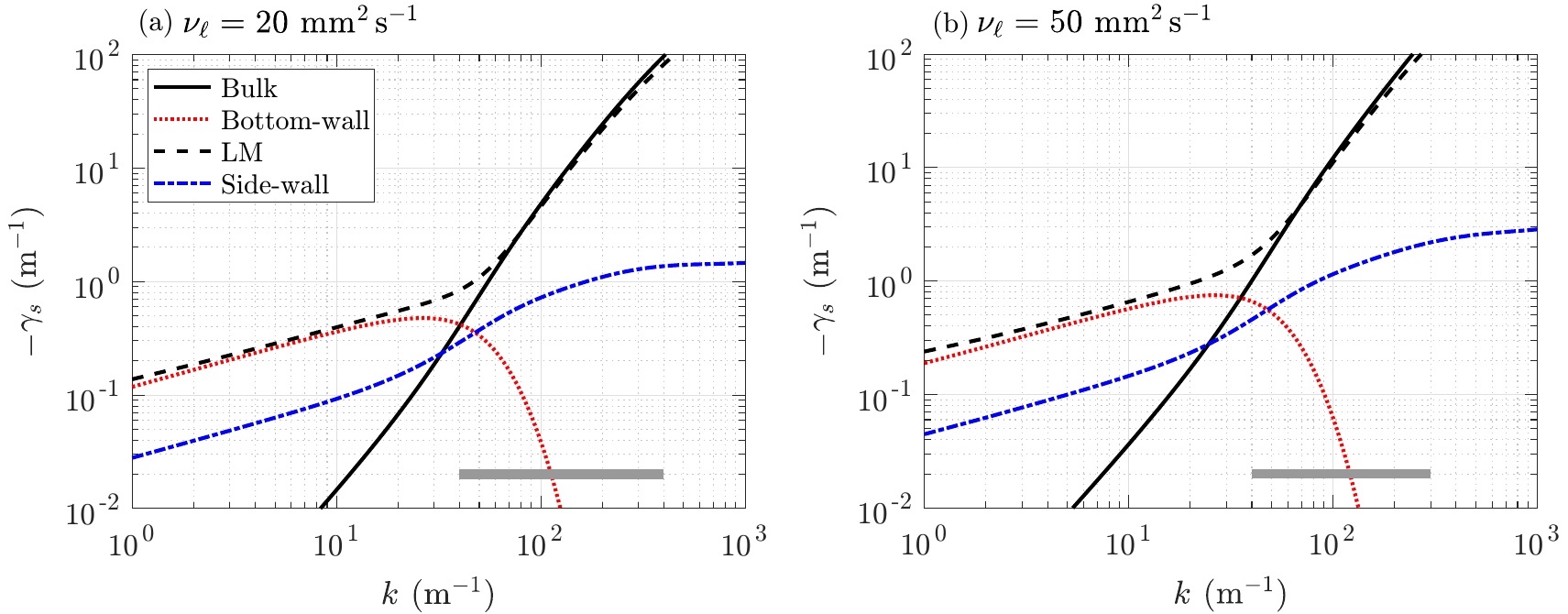}
    \caption{Various contributions to the spatial damping rates $-\gamma_s = - \gamma / c_g>0$ for (a) $\nu_\ell = 20.5$ and (b) 50.1~$\mathrm{mm^2~s^{-1}}$. LM indicates the numerically resolved Leblond-Mainardi solution, which includes the bulk dissipation and the bottom-wall dissipation. The gray horizontal bar shows the experimentally accessible range of wave numbers $k$.}
    \label{fig:gs}
    \end{center}
\end{figure}

To estimate which contributions dominate in our experiments (tank depth $h=35$~mm, width $W=296$~mm), we plot in Fig.~\ref{fig:gs} the spatial damping rates $-\gamma_s = -\gamma / c_g$ from the numerically resolved Leblond-Mainardi dispersion relation ($\gamma_{\rm LM}$) and from the three separate contributions ($\gamma_{\rm bulk}$, $\gamma_{\rm BW}$, $\gamma_{\rm SW}$) for the two liquid viscosities $\nu_\ell = 20.5$ and 50.1~$\mathrm{mm^2~s^{-1}}$. In the range of accessible wave numbers, shown as gray bars, the bulk dissipation $\gamma_{\rm bulk}$ (\ref{eq:QIT}) is the dominant contribution for $k >70$~m$^{-1}$, while the bottom-wall contribution $\gamma_{\rm BW}$ becomes significant for lower $k$. The side-wall contribution $\gamma_{\rm SW}$, shown by the dashed blue line, is never dominant, but it contributes approximately 20\% to the wall dissipation at low $k$.

\bibliography{Biblio_Viscous_Waves}

\end{document}